\documentclass{JAC2003}
\usepackage{graphicx}
\usepackage{amssymb}
\catcode`@=11 
\newdimen\z@ \z@=0pt 
\newskip\z@skip \z@skip=0pt plus0pt minus0pt
\def\m@th{\mathsurround=\z@}
\def\ialign{\everycr{}\tabskip\z@skip\halign} 
\def\eqalign#1{\null\,\vcenter{\openup\jot\m@th
  \ialign{\strut\hfil$\displaystyle{##}$&$\displaystyle{{}##}$\hfil
      \crcr#1\crcr}}\,}
\catcode`@=12 

\let\cl=\centerline
\def\figbox#1;#2;{\parbox{#2cm}{\epsfig{file=\figdir #1.eps,width=#2cm}}}
\def\figboxc#1;#2;{\cl{\figbox#1;#2;}\vglue-2mm}

\def\ie{{\it i.\kern-.5pt e.\kern1pt}}  
  
\def\up#1{$^{#1}$}  
\def\ifm#1{\relax\ifmmode#1\else$#1$\fi}
\def\Bbar{\ifm{\rlap{\kern.22em\raise1.9ex\hbox to.58em{\hrulefill}} B}}

\def\to{\ifm{\rightarrow}} \def\sig{\ifm{\sigma}}   
  
  \def\DAF{DA$\Phi$NE}  \def\f{\ifm{\phi}} 
 \def\pic{\ifm{\pi^+\pi^-}}  
  
\def\po{\ifm{\pi^0}}

 \def\epm{\ifm{e^+e^-}}
    
\def\Kb{\ifm{\rlap{\kern.3em\raise1.9ex\hbox to.6em{\hrulefill}} K}}

\def\noc{\relax\hglue0pt{\rlap{$C$}\raise.15ex\hbox{$\kern
.18em\backslash$}}}
\def\nop{\relax\hglue0pt{\rlap{$P$}\raise.15ex\hbox{$\kern
.18em\backslash$}}}
\def\noT{\relax\hglue0pt{\rlap{$T$}\raise.15ex\hbox{$\kern
.18em\backslash$}}}

\def\gam{\ifm{\gamma}}  
 \def\ab{\ifm{\sim}}  \def\x{\ifm{\times}}

\def\pt#1,#2,{\ifm{#1\x10^{#2}}}
  
  \def\dif{\hbox{d}}   
  
\def\bye{\end{document}}
\def\figdir{}

\begin{document}
\title{{\mathversion{bold}Measurement of $\sigma(e^+e^- \rightarrow \pi^+\pi^-)$ at $e^+e^-$ Colliders}}

\author{Juliet Lee-Franzini\\
LNF, INFN, I-00044, Frascati, Italy}

\maketitle

\begin{abstract}
\noindent At the DA$\Phi$NE-II workshop a session was devoted to the prospects of measuring the hadronic cross section at the new DA$\Phi$NE. The session included six papers, two theoretical and four experimental ones. The theory treatises, one on the muon anomaly and the other on measuring the hadronic cross section using initial state radiation at $e^+e^-$ colliders, set the background for the four experimental discussions. I summarize in the following the salient points of the session.
\end{abstract}

\section{INTRODUCTION}\noindent 
On 8 January 2004 the BNL ``$g-2$'' experiment announced that they have measured the negative muon anomaly, $a_{\mu^-}\\=\kern-3pt(g_{\mu^-}\!-\!2)/2$, to an accuracy of 0.7 parts per million (ppm), matching the precision of- and in statistical agreement with their previous measurement of the positive muon anomaly~\cite{ref:BNLexp}. As usual, these measurements once again confirm the validity of $CPT$-invariance. At the DA$\Phi$NEII workshop only the $a_{\mu^+}$ result had been available. It was known then that an apparent deviation of $a_\mu$ from the Standard Model evaluation of from one to three standard deviations exists. Unfortunately the deviation magnitude depends on whether the \epm\to hadrons cross section is used in the anomaly calculation or rather the photon spectral function derived from $\tau$ decays is used. Now four months later, despite improvements in statistical accuracies from both the $g$-2 and hadronic cross section sectors, the same discrepancies, even larger, are still here. Therefore most of the papers presented at the workshop only require some small updates in order to be still germane to the puzzles at hand.

\section{Theoretical Overview}\noindent
Eduardo de Rafael reminded us that $a_\mu\kern-3pt=\kern-3pt(g_\mu\!-\!2)/2$, and gave its - then - current experimental value $a_{\mu}^{\rm exp} =11\:659\:203(8)\!\times\!10^{-10}$. It is not out of place here to remember that $a_\mu$ is not the muon anomalous magnetic moment which is instead 2$a_\mu\mu_\mu$, where $\mu_\mu$ is the ``muonic magneton'', $e/(2m_\mu)$. I give here the new BNL result which combines both positive and negative muon measurements: $a_{\mu}^{\rm exp} =11\:659\:208(6)\!\times\!10^{-10}$ (0.5 ppm). 
He then listed the ingredients that enter into the computation of deviations of $a_{\mu}$ from zero and gave a flavor of the various theoretical approaches used for evaluating the contributions: $\delta^{\rm QED}a_\mu$, $\delta^{\rm VP}_{\rm had}a_\mu$, \kern1pt $\delta^{\rm lbl}_{\rm had}a_\mu$ and $\delta_{\rm electroweak}a_\mu$. VP and lbl stand for vacuum  polarization and light by light. The uncertainty on the leptonic QED contribution is negligible: $\delta^{\rm QED}a_\mu$=\pt11\:658\:470.35\pm0.28,-10,. The electroweak contribution is evaluated to the percent level: \pt(15.4\pm0.2),-10,. Coming to more uncertain terms, he showed that the sign of the light by light scattering contribution is unambiguously positive but assigned to its value a 50\% error: $\delta^{\rm lbl}_{\rm had}a_\mu$=\pt(8\pm4),-10,. He devoted most of his time to what still remains the most uncertain term, the hadronic vacuum polarization effects. He advocates using the analysis by HMNT which gives, including the corrected CMD2 results, $\delta^{\rm VP}_{\rm had} a_\mu$=\pt[(692.4\pm6.4)_{\rm lo}-(9.79\pm0.095)_{\rm nlo}],-10,. Adding everything: $a_{\mu}$=\pt(11\:659\:176.3\pm7.4),-10,~\cite{ref:HMNT}.
In conclusion the discrepancy between the experimental measurement and the SM evaluation is now $(32 \pm\ 10) \times 10^{-10}$, about 3$\sigma$.  

\section{$\sigma(\epm\to{\rm hadrons})$ }
\subsection{Why $\sigma({\rm had})$ ?}\noindent
Vacuum polarization corrections to the photon spectral function due to quark loops cannot be calculated at low energy from perturbative QCD. However we have the celebrated relation \cite{ref:durand}
$${\delta a^{\rm had, lo}_\mu}={1\over 4\pi^3}\int_{4m_\pi^2}^\infty \sigma_{e^+e^-\rightarrow {\rm had}}(s)K(s){\rm d}s.$$
where $K(s)\rlap{\raise1mm\hbox{\kern.8mm$\propto$}}\kern-.5mm%
\lower.8mm\hbox{\kern1.3mm$\sim$}\kern.5mm1/s$, enhancing the effect of low mass states. Some authors substitute 
$$\sigma_{e^+e^-\rightarrow\,{\rm had}}(s)\Rightarrow{4483.124\over4483.124}\;{s\over s}\;\sigma_{\,\rm had}(s)={R_{\,\rm had}\over s\times4483.124}.$$
$1/(s\times 4483.124)$ (=$4\pi\,\alpha^2/3s$) is the lowest order QED cross section for $e^+e^-$ annihilation into  massless muons.
\subsection{Measuring $\sigma(\epm\to{\rm had})$ using energy scan}\noindent
\epm\ annihilation into hadrons is dominated, below 1 GeV, by two pion production, mostly the $\rho$ meson. The cross section as a function of $s$ was originally obtained by energy scan, beginning in the late seventies, most recently at Novosibirsk. Radiative corrections are crux of the matter. Alexei Sibidanov reported on the recently published updating of the CMD-2~\cite{ref:CMD2} 1995 results, mostly due to better understanding of radiative corrections involved in the luminosity measurement and muon pair subtraction.
\subsection{$\sigma(\epm\to{\rm had})$ through Initial State Radiation}\noindent
Emission of an initial state radiation, ISR, photon of energy $E_\gamma$ lowers the energy available in \epm\ collisions from $W$ to $\sqrt{W^2-2WE_\gamma}$, making the whole range $2m_\pi\!<\!s'\!<\!s$ available for hadro-production at fixed collider $s=W^2$. To lowest order, the \gam+hadron cross section is:
$${\dif\sigma({\rm had}+\gamma)\over\dif s' \dif \cos\theta_\gamma}\!=\!{\alpha\over\pi s}\sigma_{\rm had}(s')\!\left[{s^2+s'^2\over s'(s-s')}{1\over\sin^2\theta}-{s-s'\over 2s'}\right]$$
where $s$ is the collision energy squared and $s'$ is the invariant mass of the hadronic system~\cite{ref:BKM}.
Let ${\rm had} = \pi ^+\pi ^-$, $s'=s_\pi = M^2 _{\pi^+ \pi ^-}$. We rewrite the cross section as: 
$${\dif\sigma(\pic\gam)\over\dif s_\pi\dif\cos\theta_\gamma}\!=\!{\alpha\over\pi s_\pi}\sigma(\pi\pi,\!s_\pi)\!\left[{s^2+s_\pi^2\over s(s\!-\!s_\pi)\sin^2\theta}\!-\!{s-s_\pi\over 2s}\right]$$
The result diverges at $\theta$=0, just like \sig(\epm\to\gam\gam).
For $\bar\theta<\theta_\gamma<180-\bar\theta$, \kern1mm $x=\cos\bar\theta$
$$\eqalign{&
{d\sigma(\pi\pi\gamma)\over d s_\pi}=\cr
&{\alpha\over s\pi}\left[{s^2+s_\pi^2\over s(s-s_\pi)}\:
\log{1+x\over1-x}-{s-s_\pi\over s}\;x\right]\sigma(\pi\pi,\:s_\pi)\cr
&=H(\bar\theta,s)\times\sigma(\pi\pi,\\ s_\pi)\cr}
$$
The above defines $H$, the radiator function that is dependent on $s$ as well as on the photon angular acceptance interval. The $\pi\pi\gamma$ cross section factorizes into two components:
the $H$ function, a pure QED vertex which can be computed to high precision and $\sigma(\pi\pi,\:s_\pi)$, the desired hadronic cross section whose integral is a dominant contribution to $a^{\rm VP}_{\rm had}$. Factorizability is in principle lost at higher order. In fact, interference between ISR and final state radiation (FSR) vanishes for $C$-symmetric integration and the \pic\gam\ cross section factorizes. Radiative corrections, soft photon and vertex terms, remove the divergence.

At low energy, where the hadronic cross section is dominated by the \pic\ channel, the photon does not have to be observed, and $s_\pi$ is the invariant mass of the dipion. The measured \dif\sig(\pic\gam)/\dif$s_\pi$ cross section divided by the $H$ function, appropriately computed using the same angular acceptance, gives the result, all done!. The advantages of doing it this way are many:
\begin{enumerate}
\item One does not need to operate the collider at different energies. Machines are very ornery and do not like to be treated like that at all.
\item The overall energy scale, at least in a detector like KLOE is established at $W=m_\phi$ to a very high accuracy and applies to all values of $M$(had).
\item The luminosity is measured at fixed energy, for the entire data set at once. Many, energy dependent, very painful corrections are thus avoided.
\end{enumerate}
There are however drawbacks that need to be overcome. First, final state radiation, FSR, where the $e^+e^-$ annihilation produces a photon together with the hadrons in the final state (note that now $k_\gamma^2=s_\gamma$ is not equal to $s_\pi$), is a background of the same order, counting powers of $e$, as the cross section of interest. One must perform an absolute measurement of a cross section that is only a tiny fraction of the total cross section. Specifically at DA$\Phi$NE in KLOE running at about 1019 MeV, $\sigma_{\rm Bhabha}\ab100\ \mu$barn, $\sigma_{\rm had}\ab3\ \mu$barn while $\sigma _{\pi^+\pi^-\gamma}\ab0.02\ \mu$barn. There is even a large amount of \pic\po\ states, mostly due to \f\to$\rho^{\pm,0}\pi^{\mp,0}$ corresponding to \ab0.45 $\mu$barn.
Furthermore, everything stated previously assumes that one knows {\it a priori} that the photon selected in the $\pi\pi\gamma$ event is the ISR photon. In fact, since ISR and FSR photons cannot be distinguished at all, one needs precise calculations to estimate the admixture contained in a particular experimental configuration. In addition, our illustration neglected higher order terms that means in fact  that the factorization mentioned is not total. Finally, one must be careful about FSR and multi-photon radiation because pion radiation must to be included~\cite{ref:JLF} in the integral of over $\sig _{\rm had}$ to obtain $\delta a_\mu$.

The Karlsruhe theory group of K\"uhn and collaborators has been studying these problems in the recent years and has produced a series of Monte Carlo Programs named from EVA and more recently PHOKHARA 1, 2, 3, 4\dots\ that are indispensable to the experimentalists. EVA in fact already included ISR terms to NLO, although was only to LO for the FSR. The second theory paper presented at the DA$\Phi$NE-II workshop was by Czyz of this group. He discussed specifically the application of the their Monte Carlo programs to extract the hadronic cross section at  $e^+e^-$ colliders such as DA$\Phi$NE where the two pion cross section dominates and PEP-II or KEKB at $W$\ab10 GeV, mostly at the $\Upsilon$ mesons. He reminded us that at DA$\Phi$NE one can significantly enhance ISR over FSR by accepting only photons at small angles. Of course one does this at a cost because, for low $M_{\pic}$, the opening angle in the lab of the two pions recoiling against a photon is small. Therefore small photon angle result in loss of the pions also, \ie\ no sensitivity to \pic\ production for $M_{\pic}<600$ MeV. At higher energies such as at BaBar choosing large angle photons still retains separation between the FSR photon and the hadrons, at least at moderate multiplicities. He also showed the results from the last PHOKHARA(4) that evaluates the relative amount of ISR and FSR contributions to $\sigma_{\pi\pi\gamma}$ in the KLOE configuration and included all theory references in his write up for these proceedings~\cite{ref:CZYZ}.

\section{Measuring $\sigma(\pi\pi)$ using KLOE }\noindent 
\subsection {2001 Data} KLOE is the first experiment that uses measured $\sigma_{\pi\pi\gamma}$ to extract $\sigma_{\pi\pi}$ at an $e^+e^-$ collider, while sitting at a fixed energy of about 1019.4 MeV and taking advantage of ISR to vary $s_\pi$~\cite{ref:JLF}. At the DAFNEII Workshop, the small angle photon measurement, i.e. from photons confined within forward and backward cones of 15 degrees half angle, was described, and its results were presented by Stefan Mueller~\cite{ref:SMUELLER}. At that time however, the FSR study using PHOKHARA(4) was just being implemented into the KLOE GEANT simulation, since the generator had barely been released by its authors. In Stefan's write up the systematic error due to FSR had been quoted simply from its maximal possible contribution estimated by the generator's authors. I update here the systematic errors from the completed KLOE study. The contribution to $\delta^{\rm had}a_\mu$ in the region where KLOE overlaps with CMD-2, i.e. for $0.37<s_\pi <0.93$ is: 
$$\eqalign{\delta a_\mu^{\rm had}&\kern-3pt=376.5\pm0.8_{\rm stat}%
\pm4.51_{\rm sys}\pm 3.76_{\rm theory}\kern3mm {\rm KLOE}\cr
\delta a_\mu^{\rm had}&=378.6\pm2.7_{\rm stat}\pm2.3_{\rm syst+theory}%
\kern7mm{\rm CMD-2}\cr}$$
Two conclusions can be drawn from the above comparison:
\begin{enumerate}
\item Since the two experiments are done with totally different methods and are of equal precision, the agreement between them supports the SM evaluation of the $a_\mu$ using  CMD--2 $e^+e^-$ data.
\item It confirms the $e^+e^-$ versus $\tau$ puzzle. Incidentally, the KLOE pion form factor is also indistinguishable from that of CMD--2.
\end{enumerate}
\subsection{2002 Data}\noindent
The above KLOE result are based on data that KLOE collected in 2001, some 147 pb\up{-1}. About twice as much data was collected in 2002, with superior running conditions. When one notes that the statistical error above is already minute, \ab0.2\%, it is clear that emphasis should be placed on what is necessary to bring the systematic and theory errors down to a similar level. On the experimental side, the dominant systematic errors are due to the uncertainty in the vertex efficiency, 0.7\% and the efficiency of the initial processing filter 0.6\%, both required because of high machine background levels in 2001. Background was significantly lower in 2002. On the theory side, the errors used are just those claimed by the authors of the MC generators, for ex. the luminosity error of 0.6\%, the radiator and FSR resummation errors, both 0.5\%, more a guess than rigorous derivations. We hope to make auxiliary measurements such as $\sigma(e^+e^-\to\mu\mu)$ to check luminosity, radiator and FSR validity. Forward and backward asymmetry of the pions to can test the validity of point pion assumption in radiation emission. Perhaps by giving such new inputs to the theorists they can reduce their estimate of their uncertainties. 

In addition, by using events with detected photons we will access the remaining part of the two pion cross section down to threshold. It should not be forgotten that, while the \epm\to\pic\ cross section has been remeasured above \ab600 MeV, the energy range $2m_\pi\le M_{\pic}\lesssim600$ MeV contributes about \pt100,-10, to $a_\mu$. The data used in the evaluation are however limited and uncertain. Use of $\tau$ information is clearly not advisable. Nor is in my opinion the use of the pion rms charge radius.~\cite{ref:hdez}

\subsection{Measuring $\sigma (\pi\pi)$ using KLOE at DAFNE2}\noindent
The last two experimental talks in this session were devoted primarily to future measurements. The first was by Achim Denig who addressed the topic of using KLOE to measure the hadronic cross section at an upgraded DA$\Phi$NE ~\cite{ref:DENIG}. The second was by Evgeni Solodov who gave a progress report on measuring this cross section using BaBar while sitting on the $\Upsilon(4S)$ energy. Solodov did not submit a written version of his presentation that, in fact, had the same content as that given by Michel Davier elsewhere ~\cite{ref:DAVIER}. Suffice it to say that they use radiative return and the PHOKHARA generators. Results will be forthcoming soon, in the sense that the data have been collected and they are at the analysis stage.

In contrast, Achim's discussion deals with taking data with an existing detector but at an as yet to be constructed accelerator DA$\Phi$NE-II, a highly luminous DA$\Phi$NE or at its fall back option, should the former prove unfeasible. He makes the argument that in the energy region between 1-2 GeV the hadronic cross section is relatively poorly measured and that the \pic\ channel is still important here and that the 4 pion final state becomes important and discusses how KLOE will be able to significantly improve the situation.

\section{Conclusions}\noindent
A long time ago, in 1947, Kusch and Foley, who later were my teachers in Columbia, performed the ``high precision'' and first measurement of the electron gyromagnetic factor $g$. Volume 73 (1948) of the Physical Review carries the second measurement of the $g$-value on p. 412 and on p. 416 Schwinger's statement of what it ought to be in QED.\footnote{Kusch got the Nobel prize (1955) for the electron $g$ factor measurement, Schwinger (1965) for QED.} The results were:
$$\eqalign{g_e&=2(1+0.00119(5))\cr
&\kern3cm\hbox{P. Kusch and H.M. Foley, 1947.}\cr
g_e&=2(1+\alpha/2\pi)=2(1+0.001162(1))\cr
&\kern3cm\hbox{J. Schwinger, 1947.}\cr}$$
Notice how in those days the estimate was \ab50 times more accurate than the measurement (Schwinger did not give any uncertainty, but he knew very well what it was). The important thing is however the fact that difficult, precise  measurements were instrumental in pushing theory to be able to confront them, in this case a fantastic confirmation of the then new QED.

We are today in a somewhat embarrassing situation, the muon anomaly is being measured with improving accuracy but nobody can compute the corrections due to quarks. One can get around contributions to the photon spectral function if measurements of \sig(\epm\to had) are available. It is very unlikely that better measurements will ever be made by varying the collider energy. Varying the collision energy is free in the real word, thanks to radiation and is the way to go. Already now, the statistical accuracy is at the 0.08\% for some very relevant piece of the \ab700 hadronic contribution to $a_\mu$. The full error however is still 1.6\%. The interpretation of the measurement is difficult because of incomplete QED calculation and also because of experimental uncertainties, the KLOE attempt being the first time ever these measurements were tried. With some more effort in this endeavor, it is conceivable that accuracies of $\ll1\%$ could ultimately be achieved. Certainly at \DAF\ and better any improved \DAF\ it is possible to achieve the accuracy mentioned and also perform all auxiliary measurements necessary to acquire confidence that the QED calculation are satisfactory. Help in the in the few GeV region will also come from BaBar and Belle.

I very much hope that the continued experimental efforts will spur calculations of radiative corrections to an equivalent level accuracy. This is necessary both for the Bhabha cross section and the extraction of the hadronic \epm\ annihilation cross section form annihilations into photons plus hadron. If we continue in this way, and progress is also made in computing light-by-light scattering contributions (lattice?), then the error on the prediction for $a_\mu$ in the strictest SM sense could be largely reduced, leading maybe to the glimpse necessary to step beyond the SM.


\begin{thebibliography}{99}   

\bibitem{ref:BNLexp}
G.W.Bennett et al. arXiv:hep-ex/0401008.
\bibitem{ref:HMNT}
K. Hagiwara, A.D. Martin, D. Nomura, and T. Teubner, arXiv:hep-ph/0312250.
\bibitem{ref:durand}
L. Durand, Phys. Rev. {\bf 128}, 441 (1962); M. Gourdin and E. de
Rafael, Nucl. Phys. {\bf B10}, 667 (1696)
\bibitem{ref:CMD2}
R.R. Akhmetshin et al., arXiv:hep-ex/0308008.
\bibitem{ref:BKM}
S. Binner, J.H. Kuehn and K. Melnikov, Phys. Lett. {\bf B 459} (1999) 279.
\bibitem{ref:JLF}
Juliet Lee-Franzini, Lepton Moments International Symposium 2003 Lecture: Cross Section of $e^+ e^-$ to hadrons and photon, Cape Cod, June 2003.
http://www.lnf.infn.it/~juliet .
\bibitem{ref:CZYZ}
Henryk Czyz, these proceedings, hep-ph/0402030.
\bibitem{ref:SMUELLER}
Stefan Mueller, these proceedings, hep-ex/0312056.
\bibitem{ref:hdez}
M.~Davier, S.~Eidelman, A.~Hocker and Z.~Zhang, Eur.\ Phys.\ J.\ C {\bf 27} (2003) 497
\bibitem{ref:DENIG}
Achim Denig, these proceedings, hep-ex/0403005.
\bibitem{ref:DAVIER}
Michael Davier, Progress on R Measurement through ISR with BaBar, hep-ex/0312063.
 

\end{thebibliography}
\end{document}


\begin{figure}[htb]
\centering
\includegraphics*[width=65mm]{JACpic_mc.eps}
\caption{Layout of papers.}
\label{l2ea4-f1}
\end{figure}

\bibitem{exampl-ref}
A.N. Other, ``A Very Interesting Paper'', EPAC'96, Sitges, June 1996.
http://www.JACoW.org.

\begin{thebibliography}{9}   